# A Study of Parallel Self-Organizing Map

Li Weigang
Department of Computer Science - CIC
University of Brasilia - UnB
C.P 4466, CEP: 70919-970, Brasilia - DF, Brazil
E-mail: weigang@cic.unb.br

## ABSTRACT

A Parallel Self-Organizing Map (Parallel-SOM) is proposed to modify Kohonen's SOM in parallel computing environment. In this model, two separate layers of neurons are connected together. The number of neurons in both layers and connections between them is the product of the number of all elements of input signals and the number of possible classification of the data. With this structure the conventional repeated learning procedure is modified to learn just once. The once learning manner is more similar to human learning and memorizing activities. During training, weight updating is managed through a sequence of operations among some transformation and operation matrices. Every connection between neurons of input/output layers is considered as a independent processor. In this way, all elements of the Euclidean distance matrix and weight matrix are calculated simultaneously. The minimum distance of every line of distance matrix can be found by Grover's search algorithm. This synchronization feature improves the weight updating sequence significantly. With a typical classification example, the convergence result demonstrates efficient performance of Parallel-SOM. Theoretic analysis and proofs also show some important properties of proposed model. Especially, the paper proves that Parallel-SOM has the same convergence property as Kohonen's SOM, but the complexity of former is reduced obviously.

# 1. Introduction

"Once saw, never forgotten" is a sentence which is used to describe a human sense and learning sequence. For example, a boy glanced at a lovely girl in a party. On his way home, girl's face appears again and again during his thinking. This is a distinct feature of the human brain. Generally speaking, the brain is organized in many places in such a way that different sensory inputs are represented by topologically ordered computational maps [Hay94]. In the field of artificial neural networks (ANN), this sequence is called pattern reorganization. The boy learned the girl's image just once and recognized it latter. Some kinds of artificial neural networks can simulate this sequence by repeated learning. Among the architectures and algorithms suggested for ANN, the SOM has the special property of effectively creating spatially organized "internal representations" [Koh90]. Kohonen attempt to construct an artificial system, SOM, that can show the same behavior as boy's experience through various learning. Following Kohonen's principle of topographic map formation, the spatial location of an output neuron in the topographic map corresponds to a particular domain or feature of the input data [Koh90]. In application, SOM has been proved to be particularly successful in various pattern recognition tasks. As mentioned by Grossberg [Gros98], the conventional learning is in terms of serial processing and this slowed down the acceptance of a sampling operation that could achieve task-dependent selectivity in a parallel processing environment. So, to simulate boy's behavior through just one time's learning, is still difficult for SOM.

In this paper, a Parallel Self-Organizing Map - Parallel-SOM is proposed to show the same behavior as human learning and memorizing activities. Willshaw-von der Malsburg's SOM is reconstructed in a parallel architecture. The number of neurons in both input/output layer and connections between them is equal to the product of the number of all elements (*M*) of input signals and the number of possible classification (*P*) of the data. The weight updating is managed through a sequence of operations among some transformation and operation matrices. So the conventional repeated training procedure is modified to learn just once. Note that in parallel processing environment, the developed weight updating algorithm makes Parallel-SOM to have the same competitive learning ability and convergence property as the conventional SOM. Some other parallel implementations of SOM have been discussed [Hyo97, Man90, Ope96, Sch97, Wu91]. The manner of the learning and structure of map are different from the proposed model.

In classical computing, Parallel-SOM is even less efficient than SOM. This is due to the extra competitive operations and weight transformations of the new model. On the other hand, putting all input as the neurons of layer is almost impossible. Suppose there are signals $\boldsymbol{x}$ ($x(i) \in \boldsymbol{x}$, $i=1,2,...M$); one input neuron and $P$ output neurons are needed by using SOM, but $M \times P$ input and output neurons are needed in Parallel-SOM.

In quantum computing, the unique characteristics of quantum theory may be used to represent information when the number of neurons is exponential capacity [Ven98b]. Using quantum representation $\boldsymbol{x}(i)$, $i = 1, ..., M$, the number of neurons is exponentially reduced to $Log_2 M$. When $M = 1000000$ and $P = 100$, in conventional computing, $M \times P =$ 100 millions neurons in both input and output layer are needed to implement Parallel-SOM; in quantum computing, just 27 quantum neurons are needed. With the synchronization feature of Parallel-SOM in quantum computing, the competitive operations and weight transformation will carry out simultaneously. This makes the Parallel-SOM more interesting in applications.

Since Beniof [Ben82] and Feynman [Fey82] discovered the possibility of using quantum mechanical system for reasonable computing and Deutsch [Deu85] defined the first quantum computing model, the quantum computation have been developed as a interesting multidiscipline. Specially in recent years, the appearances of Shor's factoring algorithm [Sho94] and Grover's search algorithm [Gro96] speeded up the development in this area. As an index of quantum computation study situation, a statistical result of the numbers of e-print paper in Quantum Physics [Lanl98] maintained by Los Alamos National Laboratory shows this tendency: 108 papers were published only in June 1998, two times more than in June 1996. There are some selected literatures [Beni82, Fey82, Deu85, Deu89, Sho94, Bar96, Gro96, Ben97, Ber97, Pre97, Sim97, Chu98, Jon98, Bir98] which can help readers to get a basic conception of quantum computation.

In the field of artificial neural networks (ANN), some pioneers introduced quantum computation into analogous discussion such as quantum associative memory, parallel learning and empirical analysis [Chr95, Men95, Zar95, Beh96, Pru96, Ven98a, Ven98b, Ven98c]. They constructed the fundation for further study of quantum computation in artificial neural networks. Eespecially, Ventura and Martinez's quantum associative memory (QuAM) has been attracted much attention in the community [Ven98b].

When comparing the quantum computation with artificial neural networks, one may find that it is necessary to modify the structure and learning manner of ANN to combine quantum parallelism. So the main purpose of this paper is to study new structure and learning algorithm of Self-Organizing Map (SOM). The paper firstly reviews the SOM and competitive learning law, specially in Kohonen's model. With the modification of Willshaw-von der Malsburg's network [vdM90], a parallel Self-Organizing Map (Parallel-SOM) and its weight updating algorithm are described in section 3. Using a typical classification example in section 4, the performance of Parallel-SOM demonstrated convergence results similar to Kohonen's model. More theoretic analysis and proofs are shown in section 5. Some interesting aspects of Parallel-SOM are studied including once learning mechanism, weight transformation, convergence of Parallel-SOM, algorithm complexity analysis and stop condition. To show the perspective of Parallel-SOM in quantum computation, a general gate array of quantum Self-Organizing Map (QuSOM) is introduced in section 6. Finally, some conclusions are summarized in the last section of this paper.

**2. Kohonen's model and learning algorithm**

Kohonen's model is particularly interesting for understanding and modeling cortical maps in the brain. The main objective of SOM is to transform an incoming signal pattern of arbitrary dimension into a one or two-dimension discrete map [Hay94]. This transformation is performed adaptively in a topological order fashion. A typical Kohonen's model consists of one presynaptic neuron and two-dimensional array of postsynaptic neurons. It's structure is shown in Figure 1. The input vector represents the set of input signals $x = [x(1), x(2), ..., x(M)]'$. The synaptic weight vector of neuron j is denoted by $w_j = [w_j(1), w_j(2), ..., w_j(M)]'$, $j=1,2,...,M$.

There are four basic steps involved in Kohonen's competitive learning algorithm: initialization, sampling, similarity and updating. They are summarized by Kohonen [Hay94, Koh90] as follows:

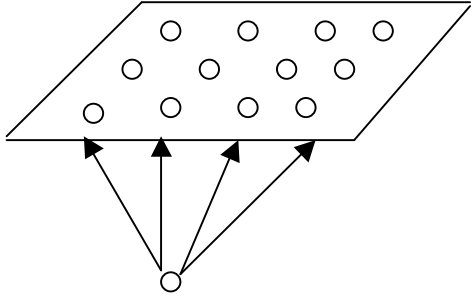

Fig. 1. Kohonen's SOM

1) *Initialization*. Choose random values for the initial weight vectors $w_j(0)$. The only restriction here is that the $w_j(0)$ must be different for $j=1,2,...,M$, where M is the number of neurons in the output layer. It may be desirable to keep small the magnitude of the weight.

2) *Sampling*. Draw a current training time sample $x(t)$, $t = 1, 2, ..., T$, from the input distribution with a certain probability; The vector $x$ $(x(i) \in x, i = 1, 2, ..., M)$, represents the sensory signal. Usually, $T > M$, and $T$ depends on the requirement of the training precision.

3) *Similarity matching*. Find the best-matching (winning) neuron $I_c(x)$ at time $t$, using the minimum-distance Euclidean criterion:

$$I_c(x(t)) = \min_j d_j(t) = \min_j \| x(t) - w_j(t) \|, \quad j = 1, 2, ..., M. \qquad (1)$$

4) *Updating*. Adjust the synaptic weight vectors of all neurons, using the update formula:

$$w_j(t+1) = w_j(t) + \eta(t)[x(t) - w_j(t)], j \in \Lambda_{I_c(x)}(t) \qquad (2)$$
$$w_j(t+1) = w_j(t), \qquad \text{otherwise}$$

where $\eta(t)$ is the learning-rate parameter, and $\Lambda_{Ic(x)}(t)$ is the neighborhood function centered around the winning neuron $I_c(x)$; both $\eta_o$ and $\Lambda_o$ vary dynamically during learning for best results. For simplicity, $\eta(t) = \eta_o [1.0 - t/T]$ and $\Lambda_{Ic(x)}(t) = \Lambda_o [1.0 - t/T]$ [Day90], where $\eta_o$ is the initial value of $\eta(t)$ and $\Lambda_o$ is the initial value of $\Lambda_{Ic(x)}(t)$.

In step 3, to find the best-matching (winning) neuron $I_c(x)$ at time $t$, $O(M-1)$ comparisons are needed. In step 4, to get a stable $w_j(t)$, the training iteration may take $O(T)$ times depending on the input distribution of $x(i)$, in many cases $T > M$. This means that the step 2 will take $\Omega(T*(M-1))$ times.

## 3. Parallel Self-Organizing Map and learning algorithm

The structure of Parallel-SOM is based on the Willshaw-von der Malsburg's model [Hay94], which consists of a two-dimensional array of presynaptic neurons and a two-dimensional array of postsynaptic. Comparing with Willshaw-von der Malsburg's model, three main differences are:

1) The number of neurons in both layers and connections between them is the product of the number of all elements ($M$) of input signals and the number of possible classification ($P$) of the data. This structure design enables one to develop once learning approach.
2) There is just one connection between an input and output neuron. Every connection is considered as a processor, the operation of every connection takes place independently.
3) The weight updating is realized through a sequence of the matrix multiplication which is a facility for parallel processing. Every element of the distance matrix and weight matrix during weight updating can be calculated simultaneously.

The structure of Parallel-SOM is shown in Figure 2.

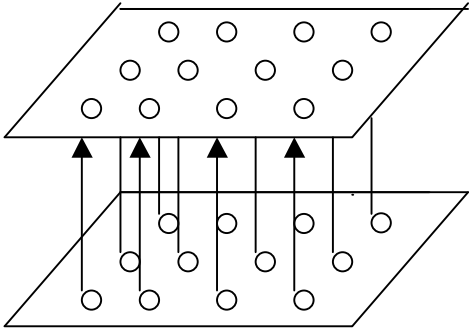

Fig. 2. The structure of Parallel-SOM

Suppose there is a series of signals $x' = (x(1), x(2), ..., x(M))$. In order to input all data once, the number of presynaptic neurons of Parallel-SOM is at least $M$. If the input data may be classified into $P$ prototypes, the structure of Parallel-SOM is designed with a two-dimensional array of presynaptic neurons $M \times P$ and a two-dimensional array of postsynaptic neurons $M \times P$. The presynaptic neurons are represented by $\mathbf{X} = (x_1, x_2, ..., x_P)$, where, $x = x_1 = x_2 = ... = x_P$. The postsynaptic neurons are represented by $\mathbf{Y}$ with the element $y(i, k)$, $i = 1, 2, ... M$, $k = 1, 2, ..., P$. Every neuron $x_k(i)$ of input layer just have one link with the neuron $y(i, k)$ of output layer with the weight $w^t(i, k) \in \mathbf{W}^t$, $i = 1, 2, ... M$, $k = 1, 2, ..., P$, where $t$ is the current operation time, $t = 0, 1, 2, ..., T$. The following is the main steps of Parallel-SOM's competitive weight updating:

1) *Once learning*: The network learns all of the input data once. Because the property of Parallel-SOM, $\mathbf{X}$ will be read in this step, where $\mathbf{X} = (x, x, ..., x)_{M \times P}$, $x' = (x(1), x(2), ..., x(M))$.

2) *Weight Initialization*. Choose random values for the initial weight vectors $\mathbf{W}^0$. The only restriction here is that $w^0(i, k) = w^0(i+1, k)$ and $w^0(i, k) \neq w^0(i, k+1)$ for $i = 1, 2, ..., M$ and $k = 1, 2, ..., P$. It may be desirable to keep the magnitude of the weight small.

3) *Similarity matching*. Repeat the steps 3,4,5,6 for $T$ times, where $M$ is called one period of weight updating, $T = n \times M$. Let $\mathbf{W}^t = \mathbf{V}$, here $\mathbf{V}$ is the production of the weight transformation matrix and weight matrix. At the first step, $\mathbf{W}^1 = \mathbf{W}^0$. Calculate all Euclidean distance $d^t(i, k)$, $i = 1, 2, ..., M$; $k = 1, 2, ..., P$, then, get distance matrix:

$$\mathbf{D}^t = \| \mathbf{X} - \mathbf{W}^t \| \tag{3}$$

and find the best-matching (winning) at time $t$, using the minimum-distance criterion,

$$d^t(i, k_{min}) = \min (d^t(i, 1), d^t(i, 2), \ldots, d^t(i, P)); \quad i = 1,2,\ldots,M. \tag{4}$$

where $d^t(i, k_{min})$ is the minimum Euclidean distance of row $i$ of $\boldsymbol{D}^t$.

4) *Updating*. Adjust the synaptic weight matrix of all neurons, by using the following update formula:

$$w^{t+1}(i, k_{min}) = w^t(i, k_{min}) + \eta(t)[\, x(i, k_{min}) - w^t(i, k_{min})], \text{ if } k = k_{min} \tag{5}$$
$$w^{t+1}(i, k) = w^t(i, k) \qquad \text{otherwise}$$

where $\eta(t)$ is the learning-rate parameter and varies dynamically during the learning for best results. For simplicity, $\eta(t) = \eta_o [1.0 - t/T]$ [Day90], where $\eta_o$ is the initial value of $\eta(t)$.

5) *Stop condition*. Verify of the condition in the following equation (6), and if (6) is satisfied then go to step 7. A precision matrix $\boldsymbol{\varepsilon}$ is simply defined by $\varepsilon(i, k) = \varepsilon$, where $\varepsilon$ is a certain small value depending on the precision requirement of the problem. There is

$$\boldsymbol{W}^{t+1} - \boldsymbol{W}^t < \boldsymbol{\varepsilon}, \tag{6}$$

6) *Reorganizing the order of matrix $\boldsymbol{W}^{t+1}$*. Multiplying the weight transformation matrix $\boldsymbol{Q}$ by weight transformation matrix $\boldsymbol{W}^{t+1}$, where $\boldsymbol{Q}\boldsymbol{Q}^{-1} = \boldsymbol{I}$. Then, a new matrix $\boldsymbol{V}$ is:

$$V = Q\,W^{t+1} = \begin{bmatrix} 0 & 0 & 0 & \ldots & 0 & 0 & 1 \\ 1 & 0 & 0 & \ldots & 0 & 0 & 0 \\ 0 & 1 & 0 & \ldots & 0 & 0 & 0 \\ & & & \ldots & & & \\ 0 & 0 & 0 & \ldots & 0 & 1 & 0 \end{bmatrix} \begin{bmatrix} w(1,1) & w(1,2) & \ldots & w(1, P) \\ w(2,1) & w(2,2) & \ldots & w(2, P) \\ w(3,1) & w(3,2) & \ldots & w(3, P) \\ & \ldots & & \\ w(M,1) & w(M,2) & \ldots & w(M, P) \end{bmatrix} = \begin{bmatrix} w(M,1) & w(M,2) & \ldots & w(M, P) \\ w(1,1) & w(1,2) & \ldots & w(1, P) \\ w(2,1) & w(2,2) & \ldots & w(2, P) \\ & \ldots & & \\ w(M\text{-}1,1) & w(M\text{-}1,2) & \ldots & w(M\text{-}1, P) \end{bmatrix}$$

(7)

7) *Registering*. Save the weight matrix $\boldsymbol{W}^{t+1}$ and stop.

## 4. Classification example

In order to compare the performance of above two SOM models, a classification example is studied. Even though this example is so simple, but it can show the work sequence of algorithms step by step. More applicable examples as satellite image classification coin counting will appear in [Wei98a, Wei98b] The data are shown in

Table 1. They are represented in Cartesian two dimension space, therefore the prototypes representing the data clusters will also be ordered in pairs [Lug98].

Table 1  Cartesian two dimensional space data

| X1 | X2 | Output |
|---|---|---|
| 1.2 | 3.0 | 1 |
| 9.4 | 6.4 | -1 |
| 2.5 | 2.1 | 1 |
| 7.9 | 8.4 | -1 |

**4.1 Kohonen's resolution**

Figure 3 shows the architecture of the Kohonen based learning network for this classification task. Two prototypes (A and B) are selected, each of them represents each one data cluster. The weight vector in node A is random initialized to (2, 4), and in node B is to (8, 6).

Usually, Kohonen learning selects data points for analysis in random order. This paper takes the point of Table 1 in the order from top to bottom for easier comparison with other method. Table 2 shows the training results. The $t$ in this table means the current training time. $(d(1))^{1/2}$, $(d(2))^{1/2}$ are the Euclidean distance. After four training iterations, the weight vector in node A converged to (2.05, 2.8), and in node B to (8.3, 7.3) which are also shown in Figure 4.

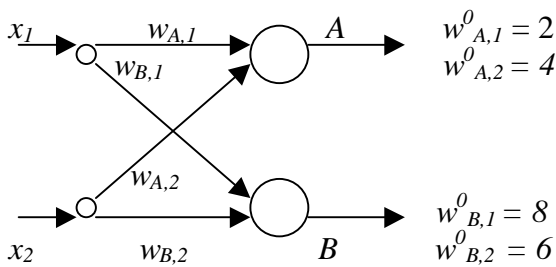

Fig. 3 Kohonen model for classification

Table 2. Training results using Kohonen's algorithm

| t | $w_{A,1}$ | $w_{A,2}$ | $w_{B,1}$ | $w_{B,2}$ | d(1) | d(2) |
|---|---|---|---|---|---|---|
| 0 | **2** | **4** | 8 | 6 | **1.64** | 55.24 |
| 1 | 1.6 | 3.5 | **8** | **6** | 69.25 | **2.12** |
| 2 | **1.6** | **3.5** | 8.7 | 6.2 | **2.77** | 55.25 |
| 3 | 2.05 | 2.8 | **8.7** | **6.2** | 65.58 | **5.48** |
| 4 | **2.05** | **2.8** | 8.3 | 7.3 | **12.04** | 122.2 |

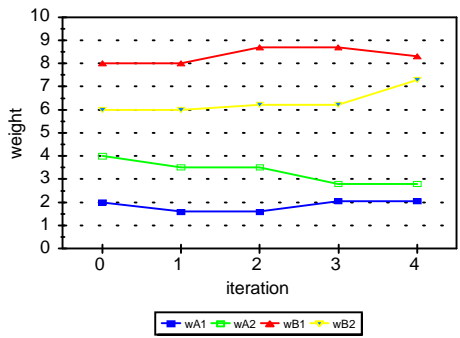

Fig. 4 The convergence of the weights using Kohonen's algorithm

### 4.2 Parallel-SOM 's resolution

In the following, the proposed model is used for two dimension data and two prototypes classification problem. However, there is a little bit difference between the following algorithm and that which was described in section 3. The data are of $N = 2$, $M = 4$ and $P = 2$. So 2x4x2 neurons are needed in both input and output layers. Figure 5 shows the distribution of input data, connections and the weights for each prototype.

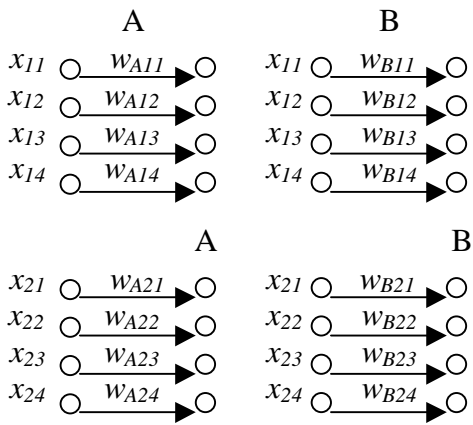

Fig. 5 Parallel-SOM model for classification

The initial weights can be randomly selected from 0 to 1 and every column of initial weight matrix can be selected to be same. In this example, they are chosen as 1, 2 for prototype A and 7, 8 for prototype B. Table 3 shows the detail sequence of the weight transformation in every weight updating step. The initial weight matrix $W^0$ is:

$$W^0 = \begin{bmatrix} 1 & 2 & 7 & 8 \\ 1 & 2 & 7 & 8 \\ 1 & 2 & 7 & 8 \\ 1 & 2 & 7 & 8 \end{bmatrix} \quad \begin{matrix} i = & 1 \\ & 2 \\ & 3 \\ & 4 \end{matrix}$$

$(d(1))^{1/2}$, $(d(2))^{1/2}$ are the Euclidean distance, and $i$ in this table means original order of the line of weight matrix. After the first weight updating iteration and $Q$ transformation, the weight matrix $V$ is

$$V = Q W^1 = \begin{bmatrix} 1.0 & 2.0 & 7.45 & 8.20 \\ 1.1 & 2.5 & 7.0 & 8.0 \\ 1.0 & 2.0 & 8.2 & 7.2 \\ 1.75 & 2.05 & 7.0 & 8.0 \end{bmatrix} \quad \begin{matrix} i = & 4 \\ & 1 \\ & 2 \\ & 3 \end{matrix}$$

After four times weight updating and $Q$ transformation, the weight matrix $V$ is

$$V = Q W^4 = \begin{bmatrix} 1.8 & 2.30 & 8.05 & 7.8 \\ 1.48 & 2.53 & 8.05 & 7.8 \\ 1.48 & 2.53 & 8.43 & 7.3 \\ 1.8 & 2.30 & 8.43 & 7.3 \end{bmatrix} \quad \begin{matrix} i = & 1 \\ & 2 \\ & 3 \\ & 4 \end{matrix}$$

the weights of first two columns of $W^4$ are close to the results of SOM (2.05, 2.8) to classify the data to the prototype A and the weights of last two columns of $W^4$ are close to the results of SOM (8.3, 7.3) to classify the data to the prototype B. The convergence tendency of elements of the first line in weight matrix is also shown in Figure 6.

Comparing figures 4 and 6, the convergence of the weights which were obtained from both models shows the same tendency. This example demonstrates the equivalence of the results in SOM and Parallel-SOM models. Further theoretic analysis of this aspect is shown in the next section.

Table 3 Training results using Parallel-SOM

| t | i | $w_{A,1}$ | $w_{A,2}$ | $w_{B,1}$ | $w_{B,2}$ | $d(1)$ | $d(2)$ |
|---|---|---|---|---|---|---|---|
| 0 | 1 | 1 | 2 | 7 | 8 | *1.04* | 58.64 |
|   | 2 | 1 | 2 | 7 | 8 | 89.92 | *8.32* |
|   | 3 | 1 | 2 | 7 | 8 | *2.26* | 55.06 |
|   | 4 | 1 | 2 | 7 | 8 | 88.57 | *0.97* |
| 1 | 4 | 1 | 2 | 7.45 | 8.2 | *1.04* | 66.10 |
|   | 1 | 1.1 | 2.5 | 7 | 8 | 84.10 | *8.32* |
|   | 2 | 1 | 2 | 8.2 | 7.2 | *2.26* | 58.5 |
|   | 3 | 1.75 | 2.05 | 7 | 8 | 78.15 | *0.97* |
| 2 | 3 | 1.75 | 2.05 | 7.45 | 8.2 | *1.21* | 66.10 |
|   | 4 | 1.1 | 2.5 | 7.45 | 8.2 | 84.1 | *7.04* |
|   | 1 | 1.1 | 2.5 | 8.2 | 7.2 | *2.12* | 58.5 |
|   | 2 | 1.75 | 2.05 | 8.2 | 7.2 | 78.15 | *1.53* |
| 3 | 2 | 1.75 | 2.05 | 8.05 | 7.8 | *1.21* | 69.96 |
|   | 3 | 1.48 | 2.53 | 7.45 | 8.2 | 77.82 | *7.04* |
|   | 4 | 1.1 | 2.50 | 8.43 | 7.3 | *2.12* | 62.15 |
|   | 1 | 1.8 | 2.3 | 8.2 | 7.2 | 74.42 | *1.53* |
| 4 | 1 | 1.8 | 2.3 | 8.05 | 7.8 | *0.85* | 69.96 |
|   | 2 | 1.48 | 2.53 | 8.05 | 7.8 | 77.82 | *3.78* |
|   | 3 | 1.48 | 2.53 | 8.43 | 7.3 | *1.23* | 62.15 |
|   | 4 | 1.8 | 2.3 | 8.43 | 7.3 | 74.42 | *1.48* |

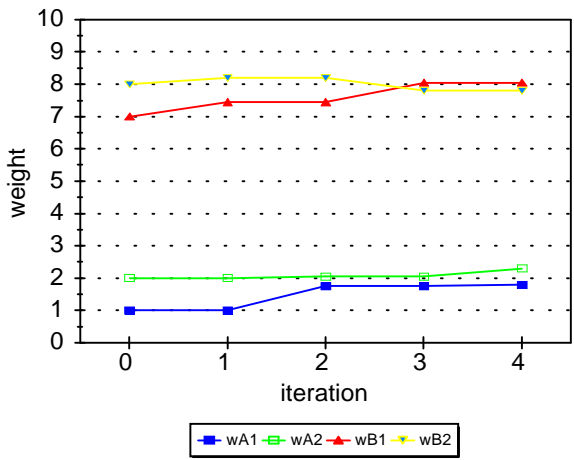

Fig. 6 Convergence of the weights using Parallel-SOM

## 5. Properties of Parallel-SOM

SOM has been received much attention in the literature because the process of it's organization is fundamental to the organization of the brain. Some intuitive principles of self-organization were summarized by von der Malsburg [vdM90, Hay94]:

1) Limitation of resources leads to competition among synapses and therefore to the selection of the most vigorously growing synapses at the expense of the others.
2) Modifications in synaptic weights tend to cooperate.

Comparing Parallel-SOM with SOM, the developed algorithm shows the satisfaction of the above principles. The detail explanation and the main properties of Parallel-SOM are described in this section.

*1) Once learning mechanism*

The learning mechanism of Parallel-SOM is different from SOM. As described in the beginning of section 3, when defining the structure of Parallel-SOM, the number of the presynaptic neurons equals the number of neurons in both layers and connections between them which is the product of the number of all elements of input signals and the number of possible classification of the data. There is just one connection between a neuron from input and a neuron from output layer. Parallel-SOM draws all samples of $x$, $(x(i) \in x, i = 1, 2, ..., M)$, just once. The competitive and weight updating is realized through a sequence of the operations which is a facility for parallel processing. So the conventional repeated learning procedure is modified to learn just once in Parallel-SOM. From this point of view, property 1 is introduced.

**Property 1**. Parallel Self-Organizing Map (Parallel-SOM) learns input signals $x$, $(x(i) \in x, i = 1, 2, ..., M)$, just once and the weights are updated through a sequence of $t$ times parallel operations.

**Proof**. The Parallel-SOM 's algorithm of section 3 can be resumed in the following sequence:
Step 1, Input $\mathbf{X} = (x, x, ..., x)$, i.e. Parallel-SOM learns all of information from outside;
Step 2, One operation of competitive and updating, for $t$ times;

2.1 $W^t = V$ (in first operation, $W^1 = W^0$);

2.2 $D^t = \| X - W^t \|$;

2.3 $d^t(i, k_{min}) = \min(d^t(i, 1), d^t(i, 2), ..., d^t(i, P))$; $i = 1,2,...,M$;

2.4 $w^{t+1}(i, k_{min}) = w^t(i, k_{min}) + \eta(t)[x(i, k_{min}) - w^t(i, k_{min})]$, if $k = k_{min}$,

$w^{t+1}(i, k) = w^t(i, k)$              *otherwise*;

2.5 If $W^{t+1} - W^t > \varepsilon$, go to 2.6, otherwise go to step 3;

2.6 $V = Q W^{t+1}$, go to 2.1;

Step 3, Saving $W^{t+1}$ and stop.

The signals *x* is input to system only at the beginning of the algorithm, at step 1, and the operations of competitive and weight updating are executed through step 2. The step 1 just passes through one time, so the property 1 is proved. At the same time, Parallel-SOM 's competitive weight updating sequence shows the satisfaction of the principle 1 of SOM.

*2) Weight transformation.*

By second principle of SOM, modifications in synaptic weights tend to cooperate. In Parallel-SOM, there is just one connection between neuron of input layer and neuron of output layer. Cooperation among neurons may be impossible when depending only on the map's structure. To satisfy this principle, weight transformation *Q* is introduced in Parallel-SOM. So the object of weight transformation *Q* is to get information from every neuron for full competition during weight updating and avoid a local minimum. This transformation will be used *T-1* times during the competitive and updating operations of Parallel-SOM. When using *Q* transformation, the position of all elements of $W^t$ will be changed after every repeated multiplication. For example, the last line of $W^t$ will become the first one and the others will be put one position backward. The table 4 shows the training results of prototypes A and B from Parallel-SOM using the data of table 1 without transformation. In this case, the minimum Euclidean distance $d_{min}$ slides down toward the direction relating to point $x_1, x_3$ for prototype A and $x_2, x_4$ for prototype B. So, any time training is no more meaning due to the local minimum. This result is also shown in figure 7: the tendency of elements of first line of weights matrix. Comparing figures 6 with

7, for prototype A, the weight updating in both cases show the convergence tendency; for prototype B, the weights keep the initial value without $Q$ transformation in figure 7. Information exchanging using weight transformation makes Parallel-SOM to have functionally the competitive learning ability and convergence property of the conventional SOM. This aspect will be proved in the next subsection.

Table 4. Training results using Parallel-SOM without transformation $Q$

| $t$ | $i$ | $w_{A,1}$ | $w_{A,2}$ | $w_{B,1}$ | $w_{B,2}$ | $d(1)$ | $d(2)$ |
|---|---|---|---|---|---|---|---|
| 0 | 1 | 1 | 2 | 7 | 8 | *1.04* | 58.64 |
|   | 2 | 1 | 2 | 7 | 8 | 89.92 | *8.32* |
|   | 3 | 1 | 2 | 7 | 8 | *2.26* | 55.06 |
|   | 4 | 1 | 2 | 7 | 8 | 88.57 | *0.97* |
| 1 | 1 | 1.1 | 2.5 | 7 | 8 | *0.26* | 58.64 |
|   | 2 | 1 | 2 | 8.2 | 7.2 | 89.92 | *2.08* |
|   | 3 | 1.75 | 2.05 | 7 | 8 | *0.56* | 55.06 |
|   | 4 | 1 | 2 | 7.45 | 8.2 | 88.57 | *0.24* |
| 2 | 1 | 1.15 | 2.75 | 7 | 8 | *0.06* | 58.64 |
|   | 2 | 1 | 2 | 8.8 | 6.8 | 89.92 | *0.52* |
|   | 3 | 2.13 | 2.5 | 7 | 8 | *0.14* | 55.06 |
|   | 4 | 1 | 2 | 7.68 | 8.3 | 88.57 | *0.06* |
| 3 | 1 | 1.18 | 2.88 | 7 | 8 | *0.02* | 58.64 |
|   | 2 | 1 | 2 | 9.1 | 6.6 | 89.92 | *0.13* |
|   | 3 | 2.31 | 2.09 | 7 | 8 | *0.04* | 55.06 |
|   | 4 | 1 | 2 | 7.79 | 8.35 | 88.57 | *0.015* |
| 4 | 1 | 1.9 | 2.94 | 7 | 8 | *0.004* | 58.64 |
|   | 2 | 1 | 2 | 9.25 | 6.5 | 89.92 | *0.03* |
|   | 3 | 2.41 | 2.09 | 7 | 8 | *0.01* | 55.06 |
|   | 4 | 1 | 2 | 7.84 | 8.38 | 88.57 | *0.004* |

*3) Convergence property*

Ritter and Schulten analyzed a Markovian algorithm for the formation of topologically correct feature maps proposed by Kohonen [Rit88] and proved that the convergence to an equilibrium map can be ensured by a criterion for the time depending on the learning step size. The following property 2 shows the general description of this convergence property of SOM.

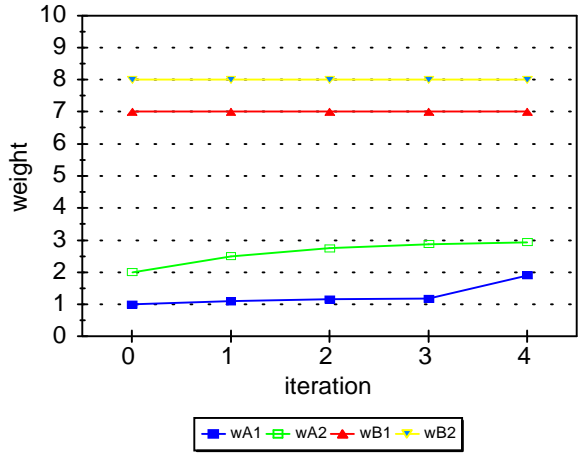

Fig. 7 The convergence of the weights
using Parallel-SOM without transformation

Let X denote a spatially continuous input (sensory) space, the topology of which is defined by the metric relationship of the vector $x \in X$. Let *A* denote a spatially discrete output space, the topology of which is endowed by arranging a set of neurons as the computation nodes of a layer. Let Φ denote a nonlinear transformation called a feature map, which maps the input space *X* onto space *A* as shown by Φ: $X \rightarrow A$. That is

**Property 2** [Hay84]. The self-organizing feature map Φ, represented by the set of synaptic weight vectors $\{w_j \mid j=1,2,...,M\}$, in the output space *A*, provides a good approximation to the input space *X*.

Based on the property 2, in the following description, it is desired to demonstrate the equivalence of Parallel-SOM and SOM in the sense of the convergence. That is

**Property 3**. Parallel Self-Organizing Map (Parallel-SOM) has the same convergence property as Kohonen's Self-Organizing Map (SOM).

**Proof.** Without loss of generality, take a simple SOM as *a)* of figure 8 and a Parallel-SOM as *b)* of figure 8 where the number of elements of input *M=3* and the number of the classification of the data *P=3*, since it can be easily generalized to arbitrary *M* and *P* situation. To verify the convergence of the map during weight updating, weights are denoted by $w^t(i, k)$, $i, = 1,2,...M$ and $k = 1,2,...,P$, for SOM and by $w^t_p(i, k) \in W^t$ for Parallel-SOM. Considering $t = n*M$, *n* is a certain integer number.

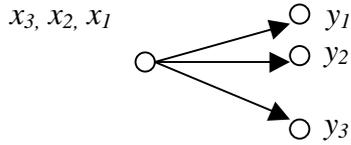

$x_3, x_2, x_1$    $y_1$
           $y_2$
           $y_3$

a) A simple SOM

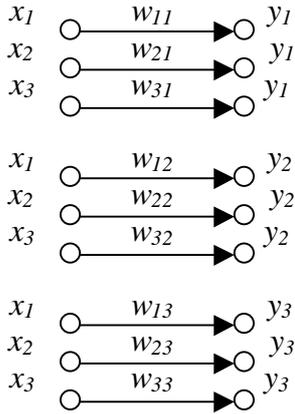

b) A simple Parallel-SOM

Fig. 8 A simple a) SOM and b) Parallel-SOM

The object of the following proof is to prove that $w^{t+M}(i, k)$ of SOM equals $w^{t+M}_p(i, k)$ of Parallel-SOM.

Consider SOM. Suppose at $t$, the input is $x(1)$ and the following sequence of input is $x(2)$, $x(3)$ and $x(1)$ (remember that each order of appearance of $x(i)$ is of probability of 1/3). From equation (2) in the case of a) of figure 8, there is

$$w^{t+1} = \begin{bmatrix} w^{t+1}(2, 1) \\ w^{t+1}(2, 2) \\ w^{t+1}(2, 3) \end{bmatrix}$$

$$= \begin{bmatrix} w^t(1, 1) + \eta(t)[x(2) - w^t(1, 1)] \\ w^t(1, 2) + \eta(t)[x(2) - w^t(1, 2)] \\ w^t(1, 3) + \eta(t)[x(2) - w^t(1, 3)] \end{bmatrix} \quad \begin{matrix} (8.1) \\ (8.2) \\ (8.3) \end{matrix}$$

$$w^{t+2} = \begin{bmatrix} w^{t+2}(3, 1) \\ w^{t+2}(3, 2) \\ w^{t+2}(3, 3) \end{bmatrix}$$

$$= \begin{bmatrix} w^{t+1}(2,1)+\eta(t+1)[x(3)-w^{t+1}(2,1)] \\ w^{t+1}(2,2)+\eta(t+1)[x(3)-w^{t+1}(2,2)] \\ w^{t+1}(2,3)+\eta(t+1)[x(3)-w^{t+1}(2,3)] \end{bmatrix} \quad \begin{array}{c}(9.1)\\(9.2)\\(9.3)\end{array}$$

$$w^{t+3} = \begin{bmatrix} w^{t+3}(1,1) \\ w^{t+3}(1,2) \\ w^{t+3}(1,3) \end{bmatrix}$$

$$= \begin{bmatrix} w^{t+2}(3,1)+\eta(t+2)[x(1)-w^{t+2}(3,1)] \\ w^{t+2}(3,2)+\eta(t+2)[x(1)-w^{t+2}(3,2)] \\ w^{t+2}(3,3)+\eta(t+2)[x(1)-w^{t+2}(3,3)] \end{bmatrix} \quad \begin{array}{c}(10.1)\\(10.2)\\(10.3)\end{array}$$

Now consider the evaluation of equations of (8.1), (9.1) and (10.1), which is the weight for classifying the prototype 1.

$$w^{t+1}(2, 1) = w^t(1, 1) + \eta(t)[x(2) - w^t(1, 1)] \quad (11.1)$$
$$= \eta(t)x(2) + (1-\eta(t))\, w^t(1, 1)$$

$$w^{t+2}(3, 1) = w^{t+1}(2,1)+\eta(t+1)[x(3)-w^{t+1}(2,1)] \quad (11.2)$$
$$= \eta(t+1)x(3) + (1-\eta(t+1))w^{t+1}(2,1)$$
$$= \eta(t+1)x(3) + \eta(t)(1-\eta(t+1))x(2) + (1-\eta(t))(1-\eta(t+1))w^t(1, 1)$$

$$w^{t+3}(1, 1) = w^{t+2}(3,1)+\eta(t+2)[x(1)-w^{t+2}(3,1)] \quad (11.3)$$
$$= \eta(t+2)x(1) + (1-\eta(t+2))w^{t+2}(3,1)$$
$$= \eta(t+2)x(1) + \eta(t+1)(1-\eta(t+2))x(3) + \eta(t)(1-\eta(t+1))(1-\eta(t+2))x(2)$$
$$+ (1-\eta(t))(1-\eta(t+1))(1-\eta(t+2))\, w^t(1, 1)$$

Suppose, that $\eta(t) = \eta(t+1) = \eta$, $t = 0$, $w^t(1, 1) = w^0(1, 1)$, $w^0(1, 1)$ is the initial weight, there is:

$$w^3(1, 1) = \eta x(1) + \eta(1-\eta)x(3) + \eta(1-\eta)^2 x(2) + (1-\eta)^3 w^0(1, 1) \quad (12)$$

Consider Parallel-SOM. From equation (5) in case of b) of figure 8, the weight matrix at $t$ step is:

$$W^t = \begin{bmatrix} w^t_p(1,1) & w^t_p(1,2) & w^t_p(1,3) \\ w^t_p(2,1) & w^t_p(2,2) & w^t_p(2,3) \\ w^t_p(3,1) & w^t_p(3,2) & w^t_p(3,3) \end{bmatrix} \quad (13)$$

To save space, let $\tau = t+1$, and $\xi = t+2$, the next 3 steps are as follows.

$$W^{t+1} = \begin{bmatrix} w^{t+1}_p(3,1) & w^{t+1}_p(3,2) & w^{t+1}_p(3,3) \\ w^{t+1}_p(1,1) & w^{t+1}_p(1,2) & w^{t+1}_p(1,3) \\ w^{t+1}_p(2,1) & w^{t+1}_p(2,2) & w^{t+1}_p(2,3) \end{bmatrix} \quad (14)$$

$$= \begin{bmatrix} w^t_p(3,1)+\eta(t)[x(1)-w^t_p(3,1)] & w^t_p(3,2)+\eta(t)[x(1)-w^t_p(3,2)] & w^t_p(3,3)+\eta(t)[x(1)-w^t_p(3,3)] \\ w^t_p(1,1)+\eta(t)[x(2)-w^t_p(1,1)] & w^t_p(1,2)+\eta(t)[x(2)-w^t_p(1,2)] & w^t_p(1,3)+\eta(t)[x(2)-w^t_p(1,3)] \\ w^t_p(2,1)+\eta(t)[x(3)-w^t_p(2,1)] & w^t_p(2,2)+\eta(t)[x(3)-w^t_p(2,2)] & w^t_p(2,3)+\eta(t)[x(3)-w^t_p(2,3)] \end{bmatrix}$$

$$W^{t+2} = \begin{bmatrix} w^{t+2}_p(2,1) & w^{t+2}_p(2,2) & w^{t+2}_p(2,3) \\ w^{t+2}_p(3,1) & w^{t+2}_p(3,2) & w^{t+2}_p(3,3) \\ w^{t+2}_p(1,1) & w^{t+2}_p(1,2) & w^{t+2}_p(1,3) \end{bmatrix} = \tag{15}$$

$$\begin{bmatrix} w^\tau_p(2,1)+\eta(\tau)[x(1)-w^\tau_p(2,1)] & w^\tau_p(2,2)+\eta(\tau)[x(1)-w^\tau_p(2,2)] & w^\tau_p(2,3)+\eta(\tau)[x(1)-w^\tau_p(2,3)] \\ w^\tau_p(3,1)+\eta(\tau)[x(2)-w^\tau_p(3,1)] & w^\tau_p(3,2)+\eta(\tau)[x(2)-w^\tau_p(3,2)] & w^\tau_p(3,3)+\eta(\tau)[x(2)-w^\tau_p(3,3)] \\ w^\tau_p(1,1)+\eta(\tau)[x(3)-w^\tau_p(1,1)] & w^\tau_p(1,2)+\eta(\tau)[x(3)-w^\tau_p(1,2)] & w^\tau_p(1,3)+\eta(\tau)[x(3)-w^\tau_p(1,3)] \end{bmatrix}$$

$$W^{t+3} = \begin{bmatrix} w^{t+3}_p(1,1) & w^{t+3}_p(1,2) & w^{t+3}_p(1,3) \\ w^{t+3}_p(2,1) & w^{t+3}_p(2,2) & w^{t+3}_p(2,3) \\ w^{t+3}_p(3,1) & w^{t+3}_p(3,2) & w^{t+3}_p(3,3) \end{bmatrix} = \tag{16}$$

$$\begin{bmatrix} w^\xi_p(1,1)+\eta(\xi)[x(1)-w^\xi_p(1,1)] & w^\xi_p(1,2)+\eta(\xi)[x(1)-w^\xi_p(1,2)] & w^\xi_p(1,3)+\eta(\xi)[x(1)-w^\xi_p(1,3)] \\ w^\xi_p(2,1)+\eta(\xi)[x(2)-w^\xi_p(2,1)] & w^\xi_p(2,2)+\eta(\xi)[x(2)-w^\xi_p(2,2)] & w^\xi_p(2,3)+\eta(\xi)[x(2)-w^\xi_p(2,3)] \\ w^\xi_p(3,1)+\eta(\xi)[x(3)-w^\xi_p(3,1)] & w^\xi_p(3,2)+\eta(\xi)[x(3)-w^\xi_p(3,2)] & w^\xi_p(3,3)+\eta(\xi)[x(3)-w^\xi_p(3,3)] \end{bmatrix}$$

Now consider the evaluation of the elements $w^{t+1}_p(1,1)$, $w^{t+2}_p(1,1)$, $w^{t+3}_p(1,1)$ from weight matrixes of equations of (14), (15) and (16), which is the weight for classifying the prototype 1.

$$w^{t+1}_p(1,1) = w^t_p(1,1)+\eta(t)[x(2) - w^t_p(1,1)] \tag{17.1}$$
$$= \eta(t)x(2) + (1-\eta(t))\, w^t_p(1,1)$$

$$w^{t+2}(1,1) = w^{t+1}_p(1,1)+\eta(t+1)[x(3) - w^{t+1}_p(1,1)] \tag{17.2}$$
$$= \eta(t+1)x(3) + (1-\eta(t+1))\, w^{t+1}_p(1,1)$$
$$= \eta(t+1)x(3) + \eta(t)(1-\eta(t+1))x(2) + (1-\eta(t))(1-\eta(t+1))\, w^t_p(1,1)$$

$$w^{t+3}_p(1,1) = w^{t+2}_p(1,1)+\eta(t+2)[x(1) - w^{t+2}_p(1,1)] \tag{17.3}$$
$$= \eta(t+2)x(1) + (1-\eta(t+2))\, w^{t+2}_p(1,1)$$
$$= \eta(t+2)x(1) + \eta(t+1)(1-\eta(t+2))x(3) + \eta(t)(1-\eta(t+1))(1-\eta(t+2))x(2)$$
$$+ (1-\eta(t))(1-\eta(t+1))(1-\eta(t+2))\, w^t_p(1,1)$$

Suppose, $\eta(t) = \eta(t+1) = \eta$, $t = 0$, $w^t_p(1,1) = w^0_p(1,1)$, $w^0_p(1,1)$ is the initial weight. There is:

$$w^3_p(1, 1) = \eta x(1) + \eta(1-\eta)x(3) + \eta(1-\eta)^2 x(2) + (1-\eta)^3 w^0_p (1, 1). \tag{18}$$

Compare the equation (12) with (18), when $w^0 (1, 1) = w^0_p (1, 1)$, i.e. the initial weights for classification of prototype 1 of SOM and Parallel-SOM are equal, then

$w^3(1, 1) = w^3_p(1, 1)$.

Continually comparing the equation (11.3) with (17.3), it is not difficult to prove

$$w^{t+3}(1, 1) = w^{t+3}_p(1, 1) \tag{19}$$

and

$$w^{t+3}(2, 2) = w^{t+3}_p(2, 2) \tag{20}$$

$$w^{t+3}(3, 3) = w^{t+3}_p(3, 3) \tag{21}$$

For the general case when the number of elements of input is $M \geq 3$ and the number of the classification of the data is $P \geq 3$, it can also be proved that

$$\begin{aligned} w^{t+M}(i, 1) &= w^{t+M}_p(i, 1) \\ w^{t+M}(i, 2) &= w^{t+M}_p(i, 2) \qquad i=1, 2, ..., M \\ &\quad ... \\ w^{t+M}(i, P) &= w^{t+M}_p(i, P) \end{aligned} \tag{22}$$

These results proved the property 3, i.e. Parallel Self-Organizing Map (Parallel-SOM) have the same convergence property as Kohonen's Self-Organizing Map (SOM).

Further analysis will show another convergent property of Parallel-SOM:

**Property 4**. For a convergent Parallel Self-Organizing Map, the elements of every column of weight matrix have a unique value to classify the data to one prototype if some input data are exactly equal to this prototype.

**Proof.** As equations (17.1), (17.2) and (17.3), one can also get the similar result for $w^{t+3}_p(2, 1)$ from equation (14), (15) and (16). Suppose, $\eta(t) = \eta(t+1) = \eta$, $t = 0$, and $w^t_p(2, 1) = w^0_p(2, 1)$, where $w^0_p(2, 1)$ is the initial weight. There is:

$$w^3_p(2, 1) = \eta x(1) + \eta(1-\eta)x(3) + \eta(1-\eta)^2 x(2) + (1-\eta)^3 w^0_p (2, 1) \tag{23}$$

Consider the condition of property 4, some input data equal the weight which classifies these data. Remember the initial weight matrix is designed as: $w^0(i, k) = w^0(i+1, k)$ and $w^0(i, k) \neq w^0(i, k+1)$ for $i=1,2,...,M$ and $k = 1,2,...,P$ in section 3. So $w^0_p (1, 1) = w^0_p (2, 1)$, and from (18), there is:

$w^3_p(1, 1) = w^3_p(2, 1)$,

In this way, there is: $w^3_p(1, 1) = w^3_p(2, 1) = w^3_p(3, 1)$, for $M=3$ and $t=3$. There is also:

$$w^T_p(i, k) = w^T_p(i+1, k), \text{ for } i=1,2,...,M \text{ and } k = 1,2,...,P \qquad (24)$$

where $w^t_p(i, k)$ is the weight element to classify prototype $k$, $T=nxM$.

This property can be verified in the case of coin classification using Parallel-SOM [Wei98c]. If there are six types of basic units of coin, such as 1, 5, 10, 25, 50, 100 cent, for any $M$ coins, after 200 steps of weight updating, the trained weight matrix from random initial weight matrix is:

$$W^{200} = \begin{bmatrix} 100 & 25 & 5 & 10 & 50 & 1 \\ 100 & 25 & 5 & 10 & 50 & 1 \\ & & \cdots & & & \\ 100 & 25 & 5 & 10 & 50 & 1 \end{bmatrix}$$

*4) Complexity analysis*

To analyze the time and space complexities, two asymptotic notations $O$ and $\Omega$ are frequently used [Hor78, 98]. In this section, time complexity of learning algorithms of SOM and Parallel-SOM are studied. In the sense of the training of the ANN, for the time complexity, the notation $\Omega$ should be used to represent a lower bound of the training operations of the algorithm. For example, $\Omega(T)$ means to training the network at least $T$ times. When $T$ is defined, the notation $O$ should be used for the time complexity within the learning algorithm such as the operations of the weight updating and the minimum distance finding. On the other hand, in this section, the following two indexes are used to represent the complexities of the learning algorithm: the operations of the weight updating (U) and minimum distance finding (C).

Firstly, the complexity of the classification of Cartesian two dimensional space data is analyzed based on the results of section 4. In the case of Parallel-SOM, two computing environments are considered: conventional (one processor) and quantum. In quantum computing, Grover algorithm [Gro96] is introduced into the Parallel-SOM to find the minimum distance during weight updating.

Table 5 The complexity analysis of the classification

| Model | Weight updating | Minimum distance |
|---|---|---|
| SOM | 16 | 8 |
| Parallel-SOM (one processor) | 64 | 16 |
| Parallel-SOM (quantum) | 4 | 4 |

The results in table 5 show that, the weight updating algorithm of Parallel-SOM is lesser efficient than SOM in conventional computing environments (one processor), but is more efficient than SOM in quantum computing.

Secondly, in general situation, when input signals is of $N$ vectors and every vector is of $M$ elements, and the data may be classified into $P$ prototypes, there are the following complexity analysis results.

For SOM, after $T$'s ($T>M$) training, the operations of the weight updating is

$$U = O(T \times N \times P) \tag{25}$$

and the operations to find the minimum distance is

$$C = O(T \times N \times (P-1)) \tag{26}$$

For Parallel-SOM, from distance matrix $D^t$, one needs to find a minimum distance using equation (4), i.e. $d^t(i, k_{min}) = \min (d^t(i, 1), d^t(i, 2), ..., d^t(i, P))$ for $i = 1,2,...,M$ times. In quantum computing, Grover's algorithm was developed in $P^{1/2}$ operations to find a certain value from a series of data with $P$ elements [Gro96, Dur96, Boy96, Bir98]. So the operations of equation (4) is also $P^{1/2}$. Using the parallel mechanism, to find $M$ minimum distances from matrix $D^t$ it will take place at the same time. In the same way, every element of weight matrix $W^t$ is updated by equation (5) simultaneously. Then there is property 5.

**Property 5**. When input signals is of $N$ vectors and every vector is of $M$ elements, and the data may be classified into $P$ prototypes, after $T's$ ($T>M$) updating, there are following complexity analysis results for Parallel Self-Organizing Map (Parallel-SOM):

the operations of the weight updating: $\qquad U = O(T) \tag{27}$

the operations to find the minimum distance:   $C = O(T \times P^{1/2})$ (28)

*5) Stop condition.*

In many cases, the input signal $\boldsymbol{x}$, $(x(i) \in \boldsymbol{x}, i = 1, 2, ..., M)$, has a large number of samples, i.e. $M \gg 1$. The weights of Parallel-SOM may converge within a satisfied precision after some iterations of the multiplication and operations. It is not necessary to repeat $M$ times, i.e. $T < M$. But in other cases, $M$ iterations are not enough for a small set of data. This is the main reason to introduce the stop condition of equation (6).

## 6. Perspective of Parallel-SOM in quantum computation

Depending on the computation environment and the application property, the parallel Self-Organizing Map may change the opinion of researchers from ANN field, in special case of its implementation in quantum computing. In the following a general discussion will be presented on the perspective of Parallel-SOM in quantum computation.

1) The most interesting feature of the Parallel-SOM is its parallelism property. Quantum mechanics computer can be in a superposition of states and carry out multiple operation at the same time. Figure 9 shows a diagram which represents a high-level Quantum Self-Organizing Map (QuSOM) gate array. The initial state of the register is on the left and time flows from left to right. Following the summarized Parallel-SOM algorithm in section 5, the $\boldsymbol{W}^t$ gate is a weight operator at $t$; $\boldsymbol{D}^t$ gate is a distance operator at $t$; $d^t(i, k_{min})$ is a minimum distance operator at $t$ which is a Grover searching oracle; $\boldsymbol{W}^{t+1}$ is a winner weight updating operator at $t$; $\boldsymbol{V}$ is a weight transformation operator at $t$, $\boldsymbol{v} = \boldsymbol{QW}^{t+1}$; $\vartheta$ is a observable extract information from register. QuSOM is developed by a sequence of the operations of these transformation and operation matrices. All the signals are input into the map just once and the output (weight) should converge by repeating this sequence.

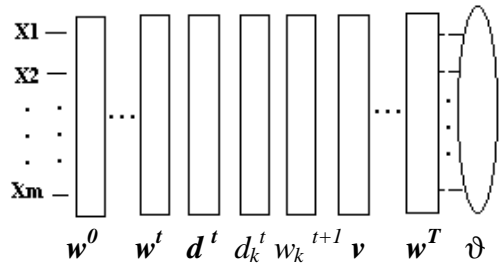

Fig. 9 A general QuSOM gate array

2) In the classical computation sense, to put all elements of the signals as neurons may be impossible and the operation of Parallel-SOM will also be time consuming. In the example in section 4, there are $N = 2$ input vectors with $M = 4$ total elements of input and $P=2$ prototypes, and the number of the neurons of both input/output layers should be 2x4x2 = 16. This number is four times the number of SOM. Fortunately, in quantum computing, this is not a problem. The unique characteristics of quantum theory may be used to represent information with a neuron number of exponential capacity [Ven98b]. For the input signals, $x(i, j)$, $i = 1, ..., M, j = 1, ... , N, k = 1, ..., P$, by using quantum representation, the neurons number is exponentially reduced to $Log_2(M \times N \times P)$. For above example, only 4 Qubits are needed for QuSOM. In some meteorological applications, the input data from satellite image may be more than 7 vectors with 43000 elements. The configuration of SOM was applied with 7 neurons of input layer and 15x15=225 neurons of output layer[Hsu96,97]. By using QuSOM, the configuration of the network may be 26 quantum input/output neurons representing 301000 neurons of SOM.

## 7. Conclusion

The study of Parallel-SOM follows the development tendency of ANN [Gros98]. The adaptation of ANN in the parallel computing environment will be interesting for both field of ANN and quantum computing, especially, for the simulation of human's learning and memorizing features by using more powerful computing tools. In Kohonen's SOM, the learning and weight updating are organized in a same sequence. This sequence is like the human's repeated learning manner. In Parallel-SOM, due to its once learning property, the weight updating is managed separately with learning and updating. This manner may appear more similarity as human's once learning way. Parallel-SOM has the same convergence property as Kohonen's SOM, but its time and space complexities are more simplified. To verify the valuation and efficiency of the algorithm, Parallel-SOM has been implemented in conventional computing (one processor) by MATLAB to meteorological satellite image classification and coin counting [Wei98a, Wei98b]. The future direction of the research is to combine Parallel-SOM with quantum computation to implement the gate array of quantum Self-Organizing Map (QuSOM) and to adapt other types of ANN into parallel computing environment.

## Acknowledgement

This research is partially supported by CNPq (Conselho Nacional de Desenvolvimento Científico e Tecnológico, Brasil) under contract no. 521442/97-4 (NV). The author thanks Dan Ventura for his valuable comments about quantum computing and Nilton C. da Silva for his help in implementing Parallel-SOM in MATLAB.